\documentclass[a4,12pt]{article}

\usepackage{mathrsfs,amsbsy,amssymb,latexsym,amsfonts,amsmath}

\catcode`\@=11
\@addtoreset{equation}{section}

\global\arraycolsep=3pt
\newcommand\path{\mathrm{path}}

\begin{document}

\begin{flushright}
{
YITP-06-38\\
OIQP-06-19\\
}
\end{flushright}
\bigskip

\begin{center}
{\Large 
Degenerate vacua from unification of second law of
thermodynamics with other laws}

\hspace{10mm}

{\large
Holger B. Nielsen}\\ 
{\it
Niels Bohr Institute,\\17, Blegdamsvej, DK2100 Copenhagen, Denmark
}
\\
and
\\
{\large
Masao Ninomiya}
{\it
\footnote{Working also at Okayama Institute for Quantum
Physics, Kyoyama-cho 1-9, Okayama City 700-0015, Japan.}\\ 
Yukawa Institute for Theoretical Physics,\\
Kyoto University, Kyoto 606-8502, Japan}
\end{center}

\vfill
\begin{abstract}
Using our recent attempt to formulate second law of
thermodynamics in a general way 
into a language with a probability density function, we
derive degenerate vacua. 
Under the assumption that many coupling
constants are effectively ``dynamical'' in the sense that they are or
can be counted as initial state conditions, we argue in our model
behind the second law that 
these coupling constants will adjust to make several vacua all
having their separate effective cosmological constants or, what is the
same, energy densities, being almost the \underline{same} value,
essentially zero.  
Such degeneracy of vacuum energy densities is what one of us works on a lot 
under the name ``The multiple point principle" (MPP).
\end{abstract}
\medskip
PACS numbers: {98.80.Cq, 95.30.Tg, 98.80.-K}
\newpage

\section{Introduction}

The second law of thermodynamics \cite{1,2,3} concerns, contrary to the other laws,
the question of initial state and further seemingly straightly
violates the time-reversal symmetry of the other laws. 
Even if time reversal symmetry is slightly broken in the Standard
Model, at last CPT is not broken, and the breaking is anyway so tiny
that it does not support the violation of time reversal invariance of
the order of that of the second law.
This arrow of time problem \cite{4} at first seems to violate any hope of
constructing a model or theory behind the second law 
without violating
the usual symmetries of the other (time development) laws, especially
CPT or time reversal symmetry. 
However, we believe to have actually presented such a model, and
S. Hawking and J. Hartle's \cite{9} no boundary initial conditions also present
a model \cite{5,6,7,8} that should indeed both have the second law for practical
purposes and obey the usual symmetries. Really our model ends up very
close to the Hartle-Hawking's one, but we think that ours is in
principle more general. 
We see the connection so that by using imaginary time
by Hawking et al have effectively got an imaginary part of the action
come in. Our model \cite{5,6,7,8,10} could be formulated as having a general complex
action where real and  imaginary parts are in principle independent
functions to be chosen only respecting the symmetries and
dimensionwise requirements etc. 

Since we ended up with a reasonable picture for second law without too
detailed assumptions about the real and imaginary parts of the action
we might claim the generalization somewhat successful.

So far we worked purely classically to avoid at first the unpleasant
quantum features of quantum mechanics for such a second
law discussion that there does not truly exist a clean history path being
true but rather a mysterious functional integral over many paths.

We did not so far go in detail with the question that such a purely
classical model could definitively not be good enough at the end.

From an esthetic and simplicity point of view it would seem that a
priori one should at first seek to construct models like the ones
mentioned, since that is what we could consider ``unification'' of the
second law with the rest of the laws and their symmetries. Also one
could easily imagine that some law behind the second law could exist
and possibly give a bit more information than just the second law
itself, so that if we could guess it or find---it is perhaps
Hartle-Hawking's no boundary---then we could use it for more. 
In our previous articles we in principle sought to discuss just a
general formulation of such a law behind the second law by simply
stating that is must---at least---be of the form of providing for
every time track---i.e. equation of motion solution---a probability
density $P(\path)$ in phase space. 
We think of the paths as associated with points in a phase space by
simply choosing a standard moment of time say $t=t_{st}$ and letting
the phase space point associated with the path be the ordered set of
generalized coordinates $q^{i}(t_\mathrm{st})$ and the ordered set of
generalized momenta $p^{i}(t_\mathrm{st})$ for this path, called path, at
that moment $t_\mathrm{st}$. 
The density $P$ shall give the probability density for the path
relative to the natural (Liouville theorem) measure on phase
space. Because of Liouville theorem saying that this measure is
invariant under the time development, the density $P(\path)$ defined
will for a given path be the same number independent of at which
moment of time $t_\mathrm{st}$ we choose to use the phase space (canonical)
density $dq dp= \prod_{i}(dq^{i}dp^{i})$. 
So generally formulated we have almost not assumed anything but left
all assumptions to be done to the selection of the functional form of
$P(\path)$ as function of the path.

At first one would think \cite{11} that $P(\path)$
should depend in a simple way only on the very first moment
$t\rightarrow 0$ or $t=t_{creation}$, the creation time of the
universe.  
However, we are with the usual law properties used as a paradigm
tempted to favor a form of the probability weight factor like
\begin{eqnarray}
P(\path)=\exp(\int P(q(t),p(t))dt)
\end{eqnarray}
 which depends in the same way on the state along the track for all
times $t$ !  
But such a form immediately seems to endanger getting out
a good second law, since its time translational invariance is already
in danger of leading to at least some features of the path to depend
is a possibly simple enough to be recognized way on even the
future. 
Such sufficiently simple dependence on the future might be
recognized as ``the hand of God'' or even ``miraculous effects" some
times. 
However, we believe that it is realistic with models of a
reasonable nature---a reasonable nice choice of $P$---of this kind to
in practice have so few miracles or ``hand of God'' effects that the
model is phenomenologically viable.  
That was what we attempted to argue for in last article \cite{10} and the
miracles would be small under the present conditions although Higgs
particles could be a special danger for them to pop up so that LHC
would be a flavored target for miracles or hand of God effects. 
The major partly future determined effect were there suggested to be
the smallness of the cosmological constant, a phenomenologically
welcome ``miracle''. 

It is the purpose of the present article to extend somewhat this
cosmological constant prediction to not only having one cosmological
constant or vacuum energy density being small, but to have several
minima in the scalar field effective potential ``landscape'' being very
close to zero, too.

This result of the present paper is what one of us (H. B. N.) and his
collaborators have been announcing as the Multiple Point
Principle\footnote{See, e.g. \cite{2} and references therein.}. 
Mainly it has been claimed to give phenomenologically good results and
derivations have, although being similar, been in principle quite
different from the present one. In fact derivations have only been
successful with some mild violation of the principle of
locality. 
In that light the previous derivations cannot be extremely
convincing, since after all, we otherwise do not find much evidence
for violation of locality, except perhaps precisely in connection with
the cosmological constant problem. 

Indeed we shall in the present article argue for the multiple point
principle, but only under a very important extra assumption: At least
some coupling constants or mass parameters are ``dynamical'', or one
should rather say that they are to be counted as part of the ``initial
conditions''. 

The meaning of this making the coupling constants---such as say the
Higgs-quark Yukawa couplings---``dynamical'' is that we consider them
part of the path in the above terminology, so that $P(\path)$ also comes
to depend on them.
Thus we have to maximize the probability also
allowing for the variation, and thus adjustment, of the couplings
which are declared \cite{12} ``dynamical''. 
We might either just assume then ``dynamical'' in this sense---really
meaning counted as part of the ``path''---as a brute force assumption,
adding them as special generalized coordinates, or we may imagine that
they in some way have come out of the ordinary dynamical variables as
e.g. in baby universe theory. 
Really it is the way of arguing in the
present article \underline {not} to go into details with respect to how precisely
the coupling constants became ``dynamical'', rather saying: 

Since we seemingly had some success---solving the cosmological
constant problem---in last articles by introducing the assumption that
the cosmological constant was ``dynamical'' in this type behind second
law model, it is by analogy suggested that also other couplings, quite
analogous to the cosmological constant, are or ``dynamical''. 

It is our hope that  allowing several possibilities for how
it came that the coupling constants became ``dynamical'' is the sense
of depending on dynamical variables or fundamentally themselves
already being ``dynamical''.
Then the model presented  has the
collected probability of being true, collected from these different
possible ways.

In the following section, section 2, we shall set up the formalism for
the probability density, and sketch how one might ideally wish it to
look very analogous with the action. 
In section 3 we shall make some rather general considerations about
the stability and most flavored states of universe that can be
relevant for surviving over exceedingly long periods of time. The main
point is here to investigate, how the likelihood of a certain
combination of macrostates $\langle Pe^{S}\rangle$ depends on the variation of the
couplings, especially when a minimum in the landscape of the scalar
field effective potential passes from being negative to being
positive. 
Our point is that the minimum being close to zero is flavored. In
section 4 we shortly review that the model could---as seen in last
article---provide an effective Big Bang although the time before the
inflation era is a crunching inflationary era with opposite second law
i.e. $ \dot{S}<0$. 
It is thus ``pre-Big Bang'' one could say. In section 5 we review how
this multiple point principle prediction has already been claimed to
be phenomenologically a very good assumption leading to
phenomenologically good predictions for relations between coupling
constants in the Standard Model, especially the top quark mass is what
is predicted.
Also a detail difference between the present and the earlier
``derivations'' of the multiple point principal of degenerate vacua is
put forward: 
In the present model many of the possible vacua are only realized over
very small space time regions. 
Perhaps only one of the vacua are hugely realized. In the old
competing derivations they all had to be realized over order of
magnitude comparable space time 4-volumes. In section 6 we present the
conclusion and further outlook. 

\section{Model behind second law of thermodynamics}

Since the second law of thermodynamics is well-known to concern the
state of the world rather than as ``the other laws'', such as Hamilton
equations or equivalently Newton's second law, then a law behind this
law must of course somehow assign probabilities to different states,
or directly tell which one is the right one. 
Since the time development laws (``the other ones'') are assumed to be
valid (under all circumstances) we should really think about a law
behind the second law of thermodynamics as assigning probability or
perhaps even validity to \underline{solutions} of the equations of
motion. We might to keep it very abstract think of a space of all
solutions to the equations of motions. 
Then the law behind the second law of thermodynamics could be
thought of as having the form of a probability distribution $P$ over
this space of solutions. 
So it (=the law behind) is required to formulate a probability measure
over this space of solutions to the equations of motion. It happens
that such a measure can be written down rather elegantly in as far as
a solution by selection of a ``standard time'' $t_\mathrm{st}$ is correlated to a
point in phase space namely 
\begin{eqnarray}
(q_{1}(t_\mathrm{st}), q_{2}(t_\mathrm{st}),..., q_{n}(t_\mathrm{st}),
p_{i}(t_\mathrm{st}),...,p_{n}(t_\mathrm{st}))~.
\end{eqnarray}
Now the phase space has the ``natural'' measure 
\begin{eqnarray}
\prod_{i}dq_{i} \prod_i dp_{i}
\label{gm}
\end{eqnarray}
which is the one from the Liouville theorem. It is of course suggested
then to use this measure with 
$(q_{1},..., q_{n}, p_{i},..., p_{n})$
taken as 
$(q_1(t_\mathrm{st}),..., q_n(t_\mathrm{st}), p_i(t_\mathrm{st}),...,p_n(t_\mathrm{st}))$ 
which means to use the measure
\begin{eqnarray}
\prod_{i}dq_{i}(t_\mathrm{st}) \prod_i dp_{i}(t_\mathrm{st}). 
\label{tspm}
\end{eqnarray}
Then one could define a density $P(\path)$ using (\ref{tspm}) by
writing the probability density for the path
\begin{eqnarray}
\path=(q_1, ..., q_n, p_1, ..., p_n) :\mbox{time axis} \rightarrow
\mbox{``Phase Space'' }
\end{eqnarray}
as 
\begin{eqnarray}
\mbox{``probability measure''} =P(\path)\prod_{i}dq_{i}(t_\mathrm{st})\cdot  \prod_{i}
dp_i(t_\mathrm{st}). 
\end{eqnarray}
One would now fear that this probability density $P(\path)$ defined this
way would depend on the standard moment $t_\mathrm{st}$ chosen. 
It is, however, trivial to see that this fear is without reason, since
indeed $P(\path)$ will \underline{not} depend on $t_\mathrm{st}$. 
It is well known that the measure
(\ref{gm}) or (\ref{tspm}) is invariant under canonical
transformations and that the time development is a canonical
transformation. Thus we do not need to attach any index $t_\mathrm{st}$ to
$P(\path)$, it is only a function of the solution ``$\path$''.

Now to really produce a guess making up a law behind the second law of
thermodynamics one has to make some assumptions about the defined
probability density function, $P:\mbox{solution space} \rightarrow
{R}_+ V\{ 0 \}$. Because if one do not assume anything it is a
very big class of possibilities for $P$ and there will not be much
content in such a formalism. 
That there is not much content in just putting up such a formalism is
encouraging, because it makes it (more) likely that we have not
assumed anything wrong by using the formalism with such $P$.

In the present article it is our intention to a large extend to keep
the model at this general level by making very  general
assumptions about $P$. 
For example we may assume that it exists for some sort of world
machinery at some fundamental level, but that we do not dare to guess
it---since our chance guessing it wrong by world of course be
outrageously high---so that we instead should attempt to guess a
statistical distribution over function of type $P$. 
Then the idea should be that we should be allowed to play with the
formalism as if $P$ were chosen as a random one from this assumed
distribution of $P$-type functions. This way of thinking of a
statistical distribution for objects---here $P$---that actually are
thought to make a law of nature is typical for the project which one
of us called ``random dynamics''. 
In this sense we can consider our last paper \cite{10} a random dynamics
derivation of the second law of thermodynamics.

Here we shall, however, not go on to put up a statistical distribution
for $P$ as a function but just keep ourselves to a rather general
discussion about $P$. 
In fact we may use such argumentation as: To find
a big value for $\log \langle P \rangle $ where $ \langle \cdots \rangle $
denotes averaging over a region in space of solutions we have less 
chance to find it very big when we average over a smaller region than
if we average over a bigger region. 
There is a bigger fluctuation for a small region and thus better
chance for the outrageous average value. 

From this kind of statistical argument we would see that it will in
all likelihood help to produce a big probability if we can get
arranged that the system would stands around in an appropriate
(not too big) region in phase space. 
The smaller this region the better is the chance that we accidentally
have in that region a high average $P$. 
Thus we would see that regions of phase which are metastable have high
chance to have some of the highest probabilities. 
As the typical region of such a stable kind or rather metastable one we could
think of the universe in a stable one of an only slightly excited vacuum
with a limited amount of field vibration on it. It might then be
metastable due to some interactions.

If we want to write down an expression for a proposal for $P(\path)$ which
has symmetry and locality properties analogous with those of the time
development laws, we would in a classical field theory model make a
construction for $\log P(\path)$ guide analogues to the action. 
The suggestion of such an analogy is in fact strongly suggested by for
a short moment thinking about a quantized generalization of our model
in a Feynman path integral formulation. 
It would  be very strongly suggested to put the $P(\path)$ in as
a factor $\sqrt{P(\path)}$ multiplying the path-amplitude by suggesting
the replacement
\begin{eqnarray}
 e^{iS[\path]}\stackrel{\mathrm{replace}}{\longrightarrow}
\sqrt{P(\path)}e^{iS[\path]}
\end{eqnarray}
(Here, $S[\path]$ is of course the action, and not the entropy.) 
 for the quality occurring in the Feynman path integral. 
To do this replacement one would of course need to have a model form
for $\sqrt{P(\path)}$ and $P(\path)$ even for those paths which do not obey the
equations of motion. 
In the present article it is, however, still the intention to use a
purely classical description and we would not need such an extension.
But only the esthetic suggestion of seeing 
\begin{eqnarray}
\log \sqrt{P(\path)}=\frac{1}{2} \log P(\path)=-\mbox{``Im$S$''}
\end{eqnarray}
as really being an imaginary part of the action, so that symmetry
and locality properties of $\log P(\path)$ would be suggested to be
taken to be just the same as for the usual---i.e. the real part
of---action $S(\path)$.
We would therefore, say in a general relativity setting, obtain a form 
\begin{eqnarray}
\log P(\path)=\int d^4 x \sqrt{g(x)} P(\varphi,
\partial_{\rho }\varphi, \psi, \partial_{\sigma } \psi, g_{\mu \nu},
\partial_{\sigma }g_{\mu\nu},\cdots) .
\label{epgr}
\end{eqnarray}
 Here we should of course have in mind that corresponding to a path one
has a development of all the field 
$\varphi(x),g_{\mu\nu}(x), \psi(x), \cdots$ their derivatives
$\partial_{\sigma} \varphi(x),\cdots$ too. 
Thus the expression (\ref{epgr}) is a well-defined functional of the
path. 

We can imagine---and it would be the most esthetic an nicest---that the
function $P$ of the fields and their derivative obey all the rules
required from the symmetries obeyed by the usual timedevelopment laws,
the ones given by the action. 

For instance since gauge transformations are supposed not to cause any
physical change, we should have $\int d^4 x \sqrt{g}$ be gauge
invariant clearly.
The form as an integral  the requirement of locality and thus
if we can manage to get such a form work phenomenologically we could
even say that the law behind the  second law of thermodynamic could
obey such a locality postulate. 

Such a set up with a lot of symmetry requirements might at first be
somewhat difficult to check and thus remain speculations, but the real
immediate worry, the reader is expected to have, is that such a form
of $P(\path)$ will have enormous difficulty in leading to the second
law. 
Immediately one would rather think that it would lead to mysterious
regularities in what will happen both in past \underline{and}
\underline{future} and even today in order to optimize $P$.
If there are too many features of the actual path predicted to be
destined to organize a special future or present the model may be
killed immediately. 

In reality we consider it a remarkable result of our previous work \cite{10}
that we argue that this type of model is \underline{not} totally out, but on the
contrary looks promising even without almost assuming anything about
the specific form of $\log P$. 

\subsection{Example:  Scalar fields, exercise}

To provide us with an idea of how such a model will function let us
imagine a theory with one or several scalar fields. 
If we add the further assumption that not only the Lagrangian density,
but also the quite analogous density $P$ has coefficients of the
dimensions required by ``renormalizability'', then the ``kinetic terms'' in
the density P would be quite analogous to the ones in the Lagrangian
density $L$ and no terms with higher number of derivatives would be
allowed neither in $L$ in $P$. 
Also only an up to fourth order term in the potential $ V(\varphi_1,
\varphi_2,\cdots)$ and the analogous ``potential'' term in $P$ would be
allowed.

To get an idea of what can go on we can think that if for some special value
combination of the scalar fields
\begin{eqnarray}
(\varphi_1,\varphi_2,\cdots)=(\varphi_1^{(0)},\varphi_2^{(0)},\cdots)
\end{eqnarray}
where the density $P$ has a maximum, then a configuration with the scalar
fields taking that set of values will be a priori very likely. 
However, we shall also have in mind that most likely the fields
will not stay at just that special combination for long if it has to
obey the equations of motion. 
Unless such a maximum in the ``potential'' part of $P$ (We think of
the part of $P$ independent of the derivatives of the fields thus only
depending on the \underline{values} of the fields) is also an extremum for the
potential part of the Lagrangian density there is no reason that
standing fields should be solutions. 
Rather the fields will roll down say---and not even any especially slow
roll a priori---.  

It might actually pay better to get a high probability or likelihood
if the field-combination chooses to sit at a minimum in the potential
$V(\varphi_1,...)$ from the usual Lagrangian density $L=\sum
\partial_{\mu} \varphi_i \partial^{\mu} \varphi_i - V (\varphi_1,...)$
with a relatively high but not maximal $P$-potential-part value.  
At such a place we could have the fields standing virtually
externally and that would count 
much more than a short stay at an even higher value for the
``potential'' part of $P$. 

The longer time of it staying there will  give much more to the
time integral form for $\log P$. 

But we can investigate if it could be arranged to get the
gain from the very high $P$ near some unstable combination for a
relatively short time and then at another earlier and/or later time
attain for long the somewhat lower but still if well-arranged
reasonably high $P$-value from a minimum in the potential $V$ from
$L$. 

In the previous articles it were suggested that such a shorter time
high $P$ could well pay and be indeed the explanation that during some
period in the middle of times there were an inflationlike Big Bang
similar time with the scalar field at an unstable point. 
Our model is not really guaranteed to solve the problem of getting the
roll slow enough---although we could say that meaning ``it would like to if it
could''--- 
but even a shorter inflation period could at least provide a from
outside (in time) seen Big Bang. 
Let us though stress two important deviations---none of which are so
far experimentally accessible---between our simulated Big Bang and the
conventional one: 
\begin{enumerate}
  \item [1)] Ours is in the ``middle of times'' so that there is a half time axis
at the pre-Big Bang side actually with an inverted second law of
thermodynamics $ \dot{S} <0$.
  \item [2)] We do not have any true singularity, but rather have inflation like
situation with finite energy density all through this ``middle period.''
\end{enumerate}

\section{The Derivation of Multiple Point Principle}
\subsection{Dynamical couplings and what to maximize }

To derive the Multiple point principle it is very important that we
take a series of coupling constants to be dynamical in the sense that
they can be adjusted to take special values guaranteeing the many
degenerate minima, which are by definition the point of the Multiple
Point Principle. 
So we must take it that the $P$-probability also depends on these
couplings. That means that so called different paths have as
some of their degrees of freedom these couplings so that they are
different for different paths. 

We have already argued for that the most likely type of path
i.e. development to corresponds to the scenario of an inflation era in
some middle of the time axis, surrounded by asymptotic  regions of an
almost static big universe with thin matter and essentially
zero cosmological constant operating near a minimum in the potential.
Then one can get the biggest $P$ from a long asymptotic
era---which though must be at least meta stable---while still getting
a high $P$ concentrated contribution from a short ``around Big Bang''
era. 

Now we should have in mind that the effective potential
$V(\varphi_1, \varphi_2,...)$ can and will typically have several
minima. 
A priori, however, these minima will not be degenerate with
their separate cosmological constants being zero as the Multiple Point
Principle which we seek to derive. 

Rather the precise height of the various minima in the effective
potential will depend on the various coupling constants and mass
parameters which we have just assumed that we shall ---at least
effectively--- count as part of the ``inial conditions'' i.e.the
solution ``path''. 
After we assumed these couplings and mass-parameters to be
``dynamical'' meaning here part of the path on which $P$ depends we
shall allow them to be varied too in the search for the most likely
path. 
Now it is, however, not quite the right thing to look for just
that very special path that goes with the highest $P$, because what we
in practice are interested in is not really to know the special path
but rather what class of paths not distinguishable by macroscopic
observation. 
We rather look for describing the scenario in terms of macrostates
meaning roughly that sort of states that are used in thermodynamics
where one characterizes systems with huge number of degree of freedom
 by means of
a few macro variables, energy, numbers of various types of particles
and the like, entropy e.g. 
Even if such a macro state having a huge number of micro states
collected under its heading does not contain the most likely single
solution to the equations of motion if it could very well happen that
the sum over all its micro states
\begin{eqnarray}
P_\mathrm{macro}=\sum_\mathrm{path-macro} P(\path)  
\label{sum}
\end{eqnarray}
could be---even much---bigger than the single $P(\path_\mathrm{max})$ for
the uttermost  scoring solution $\path_\mathrm{max}$. 
In such a case we should like in practice to consider it that the
correct scenario for us as macro-beings is the one with the macro state
giving the biggest sum (\ref{sum}). 
Rather than looking for the largest $P(\path)$ we are therefore looking
for the largest sum over a whole or perhaps even better a whole class
of similar macro states, i.e. for the largest
\begin{eqnarray}
P_\mathrm{macro}=\sum_{\path \in \mathrm{macro}} P(\path)=\langle P
\rangle_\mathrm{macro} \cdot e^S
\end{eqnarray}
where we introduced the average over the macro state notation 
\begin{eqnarray}
\langle P \rangle_\textrm{``macro''} =\frac{\sum_{\path \in \textrm{``macro''}}
P(\path)} {\mbox{\#micro  states  in  ``macro''}}=  \frac{\sum_\textrm{path
$\in$ ``macro''} P(\path)}{e^{S(\textrm{``macro''})}}
\end{eqnarray}
and defined the entropy of the macro state ``macro'' as the logarithm
of the number of micro states in it
\begin{eqnarray}
S(\mathrm{macro})\equiv \log(\mbox{\#micro  states  in  ``macro''}).
\end{eqnarray}

\subsection{Central derivation of many degenerate vacua.} 

When one characterizes the competing classes of microstates
as macrostates with some entropy $S$, 
what we really shall think of as being maximized by the model, 
is the quantity $\langle P
\rangle e^S \mu _S^{2N}$ or we can say $\log (\langle P
\rangle e^S ) $. 
Here $\langle P \rangle$ stands for the average over the macrostate of
$P$.
This quantity  $\log (\langle P \rangle L^S) $ is expected from general
smoothness assumptions and assuming no fine tuning a priori to vary
smoothly and with non-zero slope as a function of all the parameters,
especially as a function of the various coupling constants and
mass parameters. 
In other words these coupling constants and mass parameters should be
determined together with the class of microstates to be most likely
from the maximization of $\log (\langle P \rangle e^S) $ point of
view.
  
Now, however, we have to take into account that the appearance of a
minimum in the effective potential---as function of the effective
(composite or fundamental) scalar fields---in addition to that minimum
that leads the exceptionally high $\log (\langle P \rangle e^S) $ which
gives the highly probable asymptotic behavior can cause a
destabilization. 
In fact the appearance of a competing different minimum means when it
becomes deeper than the high $\log (\langle P \rangle e^S) $ one that
the latter becomes strictly speaking unstable. 
It can namely in principle then happen that the high $\log (\langle P
\rangle e^S) $ macrostate around this latter minimum, develops into a
state around the lower energy density vacuum, a state belonging 
to this other minimum. 
One should have in mind that it is the lack of energy that
keeps the ``asymptotic'' state of the universe to remain very close to
the vacuum so as to ensure the high $\log (\langle P \rangle e^S) $. 
If energy can be released by the scalar fields shifted to a
lower/deeper minimum then this cause of stability disappears and the
universe will no longer keep at the vacuum with high $\log (\langle
P \rangle e^S) $ and most likely a much lower value for $\log
(\langle P \rangle e^S) $ will be reached. That means that the smooth
continuous variation with the coupling constants etc. as a function
gets a kink, a singularity, wherever a competing minimum passes from
being above the high $\log (\langle P \rangle e^S) $ one to being deeper.

There is a very high chance that the maximum achievable $\log (\langle
P \rangle e^S) $ will occur just at this type of kink. 
All that is needed is really that as the minimum competing with the
high $\log (\langle P \rangle e^S) $ as a function of some coupling, $g$
say, is lowered---still while being above and thus no threaten to the
high $\log (\langle P \rangle e^S) $ the 
$\log (\langle P \rangle e^S)$ 
is---accidentally---having appropriate sign of its rate of
variation. 
In fact what is needed is that the $\log (\langle P \rangle e^S)
$-quantity gets larger under variation of say $g$ when the competing
minimum gets lower. 
In such a case the largest $\log (\langle P \rangle e^S) $ will 
be reached by bringing the competing minimum to be as low as possible
before it destabilizes the high $\log (\langle P \rangle e^S) $ vacuum
and thus spoils the smooth estimation. 
But that means that the maximum $\log (\langle P \rangle e^S) $
meaning the most likely scenario will precisely happen when the
destabilization sets in. 
So it is very likely that seeking---as our model does---the maximal
$\log (\langle P \rangle e^S) $ scenario will lead to very likely have
competing minima just with the same effective potential values as the
high log $\log (\langle P \rangle e^S) $---vacuum. 
But this is precisely what we mean by the multiple point principle: 
There shall be many vacua with the same energy density or
we can say same cosmological constant.

In this way out of our model we have interestingly enough derived just
this principle on which one of us and his collaborators have already
worked a lot, seeking to show that it has very good phenomenological
fitting power.

\section{Review of the other good features of our model}

In this section we shall review and elaborate the point that our
model---although it does not at first look so---indeed is to a very
good approximation a law behind the second law, even with a few extra
predictions. 

The most surprising is that we can get the second law of
thermodynamics out of an at the outset totally time reversal invariant
``law behind the second law of thermodynamics'' 
However, that can also only be done by a slight
reinterpretation: 

We argued that although the bulk of the---assumed
infinite---time axis is taken up by eras in which roughly the maximal
contribution from these bulk eras to $\log (\langle P \rangle e^S) $ is
the biggest attainable for a rather limited stable region in phase
space, it pays nevertheless to have a short less stable era in some
smaller interval.  
The full development will, in this case even if not exactly, then
with respect to crude features be time reversal invariant around a
time-reflection point in the middle of this unstable little era. 
The time reversal asymmetry is now achieved by postulating that we
ignore and in practical life do not take seriously one of the two half
axis of the time axis. 
Indeed we claim that we in practice only count what happens offer the
mentioned middle point of the relatively short ``more unstable era''. 
The argument was now that by finding some small subset of microstates
with very high $\log P$-contribution from this ``unstable'' era a
universe development with higher $\log (\langle P \rangle e^S) $
could likely be found with such an unstable period than as a
development of the type behaving as the asymptotically stable way at
all times.
Typically a very small phase space volume in the central part of the
``unstable era'' is expected to be statistically favorable because we
expected it to be easier to find an average over $P$ to be very big if we only
average over a very small region. 
We almost expect a state with exceptionally high $P$ to have to be
past to make the ``unstable era---excursion'' from the asymptotic
behavior to be the very most likely. 
We thus see that we expect the entropy in this ``unstable era''
to be very low indeed. Thinking of the especially high $P$ being
achieved by going to a highest ``potential'' part for $P$ and having
scalar fields sliding down from there the argument for a very low
entropy in the unstable era seems indeed to be justifiable in such a
more concrete setting.
 
A priori one would now think that an analogous argumentation of most
exceptionally high $\log \langle P \rangle $ occurring more likely in a
small region of phase space than in a larger phase space region would
also give a low entropy in the asymptotic era. 
Now, however, there are some phenomenological peculiarities in nature
which are combined with our suggested picture of a big universe in the
asymptotic era points to that for practical purposes $\log P$ gets almost
constant over the relevant neighborhood or the high  $\log (\langle P
\rangle e^S) $ providing vacuum (minimum in the effective potential). 
This phenomenological peculiarity is that the universe even today
already expanded so much and the parameters of the Standard Model are
such that: 
\begin{enumerate}
  \item [1)] Interactions are relatively seldom---i.e. weak couplings,
  \item [2)] All the particles around are in practice of the nature that they
only acquire non-zero-masses by the Higgs field expectation value
$\langle \varphi_{ws} \rangle \neq 0 $,
  \item [3)]  Even this Higgs VEV is tiny from the presumed fundamental scale point
of view. 
\end{enumerate}

 As a caricature we may thus see the present era---which is already
really the asymptotic era to first approximation---as an era with a
big universe with a ``gas'' of massless weakly interacting particles
only.

Further we should keep in mind that we phenomenologically have---locally at
least---Lorentz invariance. 
This means by imagining the theory rewritten from the field theory
description, used so far in this article, to a particle description
that the contributions to $\log P$ should be integrals along the time
tracks of the various particles with coefficients depending on which
particle type provides the $\log P$-contribution. 
Now, however, for massless particles the time-track is lightlike and
{\em thus always zero}. 
We get therefore no such contribution from the presumably almost
massless particles in the Standard Model. 
If this is so it means that once we have got limited the set of states
at which to find the true state in the asymptotic era to those with
the Lorentz invariance and masslessness properties, there is no gain
for $\log \langle P \rangle$ by further diminishing the class of states
included.
The $\log \langle P \rangle$ would anyway remain much the same even if
in the asymptotic time the photons say were removed because they due
to the masslessness do not count anyway. 
Thus a further reduction in phase
space in the future is not called for since really it is rather as
earlier stressed $\log (\langle P \rangle e^S) $ which should be
maximized, and we can increase \underline{this} quantity by having more
particles---meaning a wider range of phase space---contributing to
entropy $S$ without changing $\log \langle P \rangle$ much.

This masslessness phenomenology thus provides an argument for a much
higher entropy in the future in the scenario favored by our
model. Well, we should rather than future say in the numerically
asymptotically big times.

In the inflation era on the other hand the typical temperatures at
least after ``re''heating are much higher and at least the Weinberg
Salam Higgs cannot be prevented from appearing.

\section{Multiple point principle already somewhat successful
phenomenologically}

Accidentally the derivation from our law behind the second law of
thermodynamics of there being many minimal in the effective potential
for the scalar fields---fundamental or bound state ones---having all
very small cosmological constants(=potential heights) is just a
hypothesis---called multiple point principle---on which one of us and
his collaborators have worked a lot and claim a fair amount of
phenomenological success. 

In fact we started by fitting fine structure constants in model with a
bit unusual gauge group by means of the phase transition couplings in
lattice gauge theories. 
Now phase transition couplings would mean couplings for which more
than one phase of the vacuum can coexist.
So asking for vacua with the same cosmological constants is in fact
equivalent to ask for some relevant coupling constant being at the
phase transition point. 
So if we look the lattice gauge theory serious to really exist in
nature, or just the lattice artifact monopoles which mainly determine
the phase transition couplings, the above prediction of degenerate
vacua would imply such phase transition coupling constant values. 
In the old times we had indeed a sort of historically probable success
in the sense that we had the by that time unknown number of families
of leptons and quarks as a fitting parameter relating the ``family
gauge group'' gauge couplings taken to be just at the phase transition
point, and we fitted it to be three. 
Thereby we predicted by a model that had as one of its major input
assumptions the equally deep minima---although formulated rather
differently---just derived. 
The model though is just one among many possibilities first of all
characterized by having the gauge group of the Standard Model 
$G=SU(3) \times SU(2) \times U(1)=S(U(2) \times U(3))$
 repeated at a more fundamental level near the Planck energy scale
once for each family of quarks and leptons. 
In other words, each of the $N_{gen}$ families of quarks and leptons
supposed finally to be found had their own set of Standard Model gauge
particles only acting on that ``proto''family. 
Remarkably we predicted this number of families $N_{gen} \approx 3$
\underline{before} the measurement at L.E.P. of the number of families of
neutrinos. 

In the pure Standard Model our requirement of same dept in this case
of a second minimum in the Weinberg-Salam Higgs effective potential as
the one of which we live $\langle \varphi_{ws} \rangle \approx 
24 GeV/ \sqrt{2} $ leads to the Higgs particle to be the minimal one
allowed by stability of vacuum.
Without extra corrections pure renormalization group calculations lead
to a prediction of the Higgs mass from this degeneracy principle to be
$135 GeV/c^2$. 
This is already good in consideration of indirect Higgs mass
determinations pointing to a light Higgs mass. 

In works involving one of us (H.B.N.) and C.D. Froggatt and
L. Laperashvili were developed a perhaps not so trustable story of an
exceptionally strongly bound highly exotic meson of 6 top quarks and
6 antitop quarks bound together by Higgs exchange just in such a way
as to produce a degenerate vacuum with this type of exotic meson
forming a Bose-condensate.
Remarkably enough our calculations taking that sort of bound state or
exotic meson serious and imposing the degeneracy of the vacua, not
only leads to an only within uncertainly too high Yukawa coupling for
the top quark, but also solves the problem essentially behind the
hierarchy problem! 
Indeed the coincidence of the top-quark-Yukawa-coupling values $g_t$
needed for 
\begin{enumerate}
  \item[1)] getting the bound state condensate just be degenerate,
and for 
  \item[2)] getting it possible to have the second minimum in the
Weinberg-Salam Higgs field effective potential degenerate with the
first one;

\end{enumerate}
 leads to a need for the ratio of the Higgs vacuum expectation values
the minima be a number given as an exponential.

That is to say that if for some reason the second minimum in the
Weinberg-Salam Higgs effective potential were of the order of some
grand unifying scale or the Planck scale or a fundamental scale, then
the ratio of this scale to the weak scale would be explained to have
to be an exponentially big ratio from the
derived multiple point principle in the present article. 
In this sense we can claim that the
multiple point principle solved the question as to 
why so big a scale ratio problem,
a problem which is really behind the more technical hierarchy problem.

\section{Conclusion and Outlook}

We have worked further on the model that the second law of
thermodynamics be caused by there existing a ``fundamental''
probability density functional $P$ assigning to each possible solution
``path'' of the equations of motion a probability density
$P$(``$\path$'') in phase space. 
Without making more than even in mild form the assumption that this
``fundamental'' probability assignment $P$ should obey the usual
properties of laws of nature---locality (in time first of all) and
translational invariance---we already got (phenomenologically) good
results. 
In fact we roughly and practically got the second law of
thermodynamics as was the initial purpose and in addition some good
cosmology.

The present article obtained the further prediction of there being
most likely many different states of vacuum, all having small
cosmological constants. It must be admitted though that we only
obtained this result with the further very important assumption
that---some way or another the coupling constants and mass parameter,
i.e. the coefficients in the Lagrangian density, have become or are
what we call ``dynamical''. This meant that it somehow were
themselves or depended on ordinary dynamical variables, like fields or
particle positions. 
Now it turned out remarkably that this prediction by one of us and his
collaborators had since long been argued to be a good one
phenomenologically! 
It must be admitted though that for all its successes a bit of helping
assumptions were to be used.  
But even with only a mild assumption that the order of magnitude of
the Higgs field in the high Higgs VEV alternative vacuum we got a very
good value for the top quark mass $173GeV \pm 6 GeV$. 
Taking our previous Multiple Point fitting most seriously with three
degenerate vacua in the Standard Model alone we actually could claim
that Higgs-mass of $115 GeV/c^2$ seemingly found at L.E.P. is quite well matching
as being our prediction.  
It must be admitted though that especially our correction bringing the
predicted down from our older prediction $135 GeV/c^2$ to about the
L.E.P. values is very doubtful and uncertain.

It is remarkable that we get such a funny and at least in future by
Higgs mass, testable series of models higher scale of energy
predictions about couplings constants as this multiple point principle
out of modeling the second law of thermodynamics, an at first sight
rather different branch in physics. 
Already this, provided it works (i.e. Higgs mass really be
what the calculations will give etc.), would be a remarkable sort of
unification of this second law with other physics information,
seemingly at first quite unrelated! 
Taking into account that the major development of the universe into a
low density, low temperature, large universe could---in the foregoing
articles in this series---be considered the major ``hand of God
effects'' predicted from of model we must say that it unifies quite
far away features for the physical world! 

As outlook we may list a few routes of making testing of our present
unification: 

\begin{enumerate}
  \item [1)] In the light of the result of the present article testing of
there being the many degenerate vacua in the various models beyond the
Standard Model may if sufficiently successful be considered a
confirmation of our ``law behind the second law of thermodynamics''. 
  \item [2)] One could seek to estimate more numerically the cosmological
parameters such as what size the already argued to be ``small''
cosmological constant should have included here could also be if some
detail concerning the inflation going on predicted could be tested by
say microwave background investigations.
  \item [3)] A third route of testing or checking the model would be to really
find rudimentary ``hand of God effects''. 
That would of cause from the conventional theory point of view be
quite shocking and thus be a strong confirmation of something in the
direction of our model, if such effects were convincingly seen. 
It would of course be even more convincing if they were found with a
predictable order of magnitude and of the right type. 
In previous articles we put is as an especially likely possibility
that Higgs particles---special in the Standard Model by not being mass
protected---were either flavored or disfavored to be produced. That
is to say there would respectively happen hand of God effects seeking
to enhance or to diminish the number of Higgs particles being
produced.
\end{enumerate}

\section*{Acknowledgments}
We acknowledge the Niels Bohr Institute (Copenhagen) and 
Yukawa Institute for Theoretical Physics for their hospitality
extended to one of them each.
The work is supported by Grant-in-Aids for Scientific Research 
on Priority Areas,  Number of Area 763 ``Dynamics of Strings and Fields", 
from the Ministry of Education of Culture, Sports, Science and Technology, Japan.

\end{document}